\documentclass[peerreview,12pt]{IEEEtran}
\IEEEoverridecommandlockouts
\usepackage{cite}
\usepackage{amsmath,amssymb,amsfonts}
\usepackage{algorithmic}
\usepackage{graphicx}
\usepackage{textcomp}
\usepackage{xcolor}
\usepackage{setspace} 
\def\BibTeX{{\rm B\kern-.05em{\sc i\kern-.025em b}\kern-.08em
    T\kern-.1667em\lower.7ex\hbox{E}\kern-.125emX}}
\begin{document}

\title{A Novel Energy-aware Design for a D2D Underlaid UAV-Aided Cognitive Radio Network\\
{\footnotesize \textsuperscript{}}
}

\author{\IEEEauthorblockN{Hossein Mohammadi Firouzjaei,\ \ Javad Zeraatkar Moghaddam,\ \ Mehrdad Ardebilipour}}

\maketitle

\begin{abstract}
This paper investigates the effectiveness of a free disaster response network (free-DRN) and an energy harvesting enabled DRN (EH-enabled DRN) enhanced by cognitive radio (CR) technology and unmanned aerial vehicles (UAVs). In the EH-enabled DRN scenario, where device-to-device (D2D) communication harvests energy from cellular users, significant improvements are demonstrated in terms of energy efficiency (EE) and communication rate compared to the free-DRN approach. The impact of user density and UAV specifications on network performance is analyzed, addressing the challenge of optimizing the duration of energy harvesting for both cellular users and D2D devices in the EH-enabled DRN scenario. Additionally, simulation results reveal an optimal UAV height that ensures efficient network operation for varying densities of D2D devices. Overall, numerical and simulation findings highlight the superior performance of the EH-enabled DRN approach, showcasing the positive effects of enabling D2D links and improved EE. Notably, reducing energy harvesting duration and cellular user density can further enhance the EE of the DRN by up to 3dB.
\end{abstract}


\section{Introduction}
Unmanned Aerial Vehicles (UAVs), commonly known as drones, offer remarkable capabilities such as agility in flight and easy deployment, making them an ideal solution for various applications. In particular, leveraging drones for providing telecommunication services in areas affected by natural disasters, such as floods or earthquakes, holds tremendous promise [1], [2]. Additionally, scenarios like sports stadiums or multi-day festivals, requiring short-term but high-speed communication services, can greatly benefit from drones serving as telecommunication stations [3], [4]. In disaster response networks (DRNs), where ground base stations may not be accessible, UAVs acting as base stations (BSs) play a crucial role in enabling effective communication [5]. However, in DRNs, device-to-device (D2D) users often face energy constraints when establishing D2D links, necessitating the incorporation of energy harvesting (EH) techniques to support their communication needs [6].

This paper investigates a UAV-enabled DRN, as illustrated in Fig. 1, focusing on two distinct scenarios. In the first scenario, there is no cellular-to-D2D (C2D) EH link, and D2D transmitters have limited energy for data transmission. Despite the relatively short distances between D2D pairs, establishing D2D links requires energy. Furthermore, due to the absence of conventional telecommunication infrastructure, D2D users' energy resources are restricted. Consequently, this scenario exhibits weak D2D connection rates and network energy efficiency (EE) owing to the limited transmission power of D2D users.
\begin{figure}[!t]
\centerline{\includegraphics[width=2.5in,clip]{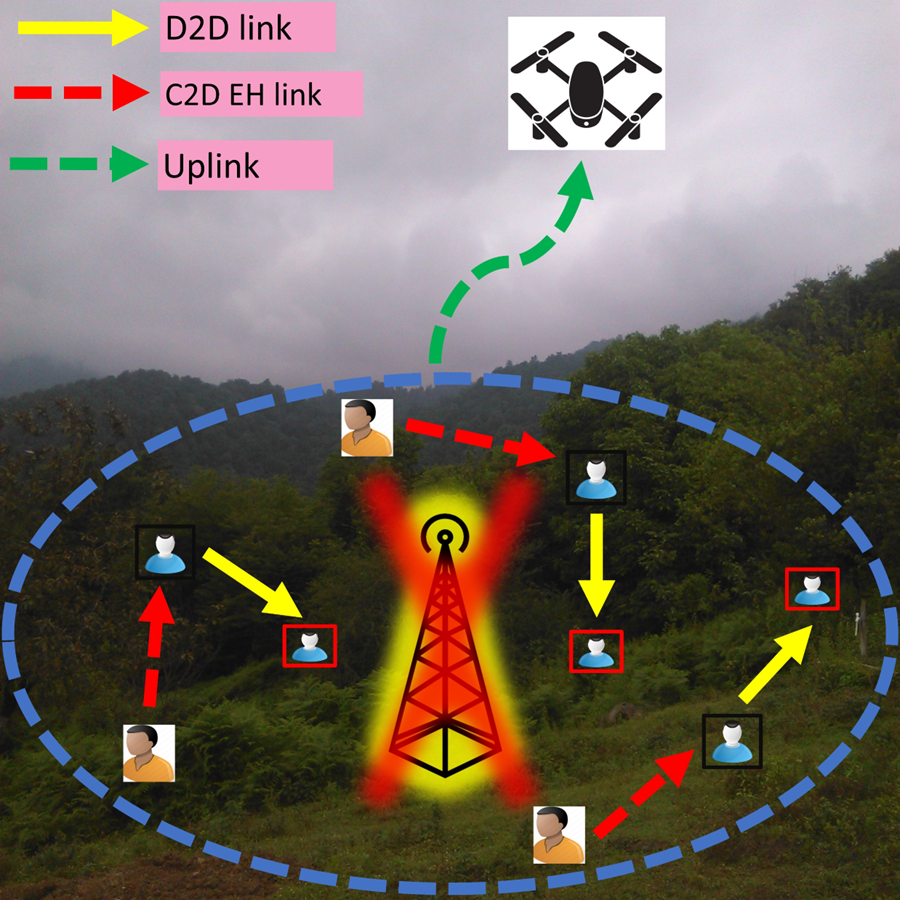}}
\caption{applying an EH technique in an UAV-enabled DRN.}
\label{fig1}
\end{figure}

In the second scenario, an EH technique between cellular users and D2D users is employed, enabling D2D users to harvest energy from cellular users. This energy infusion empowers D2D users to transmit data over greater distances, resulting in improved network EE. Moreover, D2D users, previously constrained by limited transmission energy, can achieve higher data rates.

The considered setup entails underlaid D2D users operating alongside uplink cellular users, where cellular users serve as primary users, while D2D users function as secondary users. To optimize the performance, interference management strategies are employed in both scenarios, carefully selecting the UAV altitude to accommodate varying D2D user densities. By defining and comparing the EE of the DRN in each scenario, the impact of changes in user density on EE is evaluated, highlighting the effectiveness of the second proposed scenario.

This paper further explores the influence of different network characteristics on EE, revealing that increasing the transfer power of uplink users enhances EE. Moreover, a trade-off is observed between user density and transfer power. Through comprehensive simulation results, the validity of theoretical calculations and their alignment with practical implementations is established.

The main achievements of this paper can be summarized as follows:
\begin{itemize}
  \item Enhanced EE of DRN: The introduction of an EH technique in the second scenario significantly improves the EE of an uplink network lacking ground stations. By increasing data transmission rates and reducing energy consumption, the EE improves by up to 3dB.
  \item Optimum UAV Altitude: The second scenario, leveraging cellular users' energy more effectively than the first, facilitates stronger D2D links. Consequently, the adverse effects of increasing D2D user density on uplink communication can be mitigated by operating the UAV at higher altitudes, enabling the service of more users.
\end{itemize}

By addressing these key aspects, this paper contributes to the advancement of energy-efficient and reliable telecommunication services in UAV-enabled DRNs, enabling effective communication during critical situations and dynamic event scenarios.

\subsection{Related Work}
The coexistence of uplink cellular users and D2D users along with the use of CR techniques and EH techniques to improve the rate and EE of a DRN is the main topic of this paper. Some related studies are discussed in this subsection.

In [5], EH in 5G and D2D communication is introduced as a valuable paradigm that greatly improves EE due to limited energy capacity. In the proposed scenario of [1], several UAVs simultaneously provide telecommunication service to cellular users and provide energy to D2D users. The purpose of this scenario is to increase the EE of D2D networks and reduce the communication time, and also the frequency spectrums are shared so that we can see the spectral improvement of the network along with the EE improvement. In [2], a joint beam width selection algorithm, power control, and an EH time ratio optimization algorithm are proposed for users based on alternating optimization. Iterative method is used for this problem and EE is increased to a good extent. A noteworthy point in [5] is that cellular users receive information from the UAV during the air-to-ground (A2G) channel, while D2D users harvest energy from A2G signals. Interference management is very important in [5] because cellular and D2D users interfere with each other, which greatly reduces the EE.

In [2], a framework for channel allocation for D2D, UAV deployment and power control is presented. The location of drones is optimized by repeated use of particle swarm optimization (PSO) algorithm.  After optimizing the location of UAVs, the EE of D2D communication is optimized to achieve an acceptable level of quality of service (QOS).  adaptive mutation salp swarm algorithm (AMSSA) is also proposed to solve the optimization problem [2].  In the simulation results of [2], it is shown that the frequent use of the PSO algorithm achieves better results than AMSSA algorithm. The main goal of the optimization problem of [2] is to achieve the optimal number and optimal location for UAVs in order to achieve the appropriate Signal-Noise-Interference-Ratio (SINR) value for D2D and cellular users..

In [6], the scenario of using a UAV to communicate in a disaster area where there is no safe base station is investigated.  Downlink communication is considered for cellular users and multi-hop approach for D2D communications. In such an approach to D2D communication, the drone can communicate with more users. In addition, in multi-hop D2D communication, the spectral efficiency (SE) and EE of the network are significantly improved. In [3], SE and EE have been investigated according to the change of various characteristics of the network, including the number of clusters, the distance between D2D users, and the transmission power of D2D users. The proposed scenario of [3] UAV actually plays the role of a relay between a BS and users involved in the disaster, and the clustering technique is implemented for D2D users in such a way that the UAV communicates with the head of each cluster. In [3], each cluster head receives energy from the UAV in addition to data, which is a new and significant approach.

In [7], a wireless communication system equipped with an UAV with energy harvesting is investigated. Energy is transferred in half duplex or full duplex form from the UAV as a downlink to the users.  The optimization problem of [4] minimizes the energy consumption of the UAV in such a way that the minimum data requested by the users is received and solves the path planning problem.  The energy efficiency of the UAV is calculated in the stationary state and in the moving state, and the optimal height of the UAV is obtained according to the energy consumption.  An interesting point in [4] is the optimization of the UAV flight time in the downlink connection, which is a vertical flight for a rotary wing UAV.

The uplink information transmission and downlink simultaneous wireless information and power transfer (SWIPT) in UAV-aided millimeter wave cellular networks are studied in [8]. Two various cluster processes are considered for user equipment locations in the proposed scenario of [8].The uplink and downlink phases are investigated in detail. Clustering techniques are applied in the uplink phase, where each UAV is connected to its cluster.The interesting thing that can be seen in the proposed scenario [8] is that the UAV divides the power received from users into two parts, one part for decoding the received signal and the other part for EH. In [8], to evaluate the performance of the proposed scenario, it is used to check the probability of success and energy coverage by changing the network components.

The two main challenges that UAV-aided telecommunication networks face are lack of spectrum and lack of energy. These two challenges arise from the limited energy and limited flight time of the UAV. Therefore, any network that uses UAV as a BS should include an energy efficient and spectrum efficient plan. In [9], optimization of SE and EE for UAV cognitive networks based on location information is studied. In [9], a hybrid mode is proposed, which provides high-quality communication for users based on users' location information and checking the UAV's sending power. In fact, the transmission power of the UAV is optimized to achieve a balance between SE and EE at different locations. As a result, the flight height of the drone is also optimized because the location of the UAV has a significant impact on the energy consumption of the network.

DRNs are networks that are involved in a natural disaster such as a hurricane or a human disaster such as a war, and ground base stations are out of reach [7], [8]. DRN users need a base station to communicate with each other as well as with other areas. In such a scenario, drones play a vital role and provide services to ground users (GUs). In DRNs, all terrestrial users, whether cellular users or device-to-device (D2D) users, seek to exploit the available frequency spectrum and establish a secure connection [9].  A downlink or uplink can solve the problem of cellular users to a great extent.  On the other hand, D2D users also seek to receive energy according to their specific conditions [10].

D2D users are considered as pairs, each pair consisting of a sender and a receiver. The two main characteristics of D2D communication are the limited energy of D2D transmitters to send the signal and the short distance between the transmitter and the receiver [1], [11]. The implementation of D2D communication does not require significant facilities, but there is a main challenge in D2D communication, which is the limited energy of D2D users to send signals to the corresponding receivers [3], [12]. The energy limitation of D2D transmitters reduces the probability of successful communication, thus reducing the transmission rate and energy efficiency (EE) [13]. Using energy harvesting techniques can be a promising solution to increase the probability of D2D communication success.

Various techniques for EH have been proposed in recent studies, all of which seek to increase the energy efficiency of the network [5]. In some studies, energy harvesting is in the form of downlink from UAV and for ground users, while in some models, energy harvesting is in the form of uplink and from cellular users for UAV [6].  Considering the limited energy of the UAV and the limited energy of D2D users, most of the energy harvesting techniques use cellular users as the energy source because the transmission power of cellular users is usually high.  In scenarios where cellular users and D2D users co-exist and share frequency spectrums, the importance of energy harvesting techniques is felt even more because the energy of cellular users can be used to help D2D communication [5], [14].

The coexistence of cellular users and D2D users can be examined from two perspectives, and from these two perspectives, better communication quality can be achieved. A big achievement in such a scenario is to provide energy for D2D communication, which in previous studies often had the ability to send small amounts of data [15]. In the second part, the lack of spectrum in DRNs has always been a fundamental challenge, which leads to the limitation of cellular communication and D2D communication [16]. Cognitive Radio (CR) techniques significantly increase the spectrum efficiency of the network by sharing the spectrum of cellular users for D2D users. In such scenarios, D2D users can reuse the frequency bands of cellular users when these bands are empty [2].

The CR techniques are the most important technique in the spectrum efficiency of telecommunication networks, because their main goal is the maximum use of the available frequency spectrum [17]. The lack of spectrum has always been a major challenge in telecommunications, and with the introduction of drones to telecommunication systems, the concern about the lack of frequency spectrum has increased because the drone must serve a large number of users in a limited time [4]. As a result, spectral efficiency is very important in DRN networks, especially when D2D users and cellular users coexist. In recent studies [18], a very promising method to improve spectral efficiency has been presented in such a way that by using CR in DRN networks, cellular users have the first priority in using the spectrum, and then D2D users use empty frequency bands [19], [20].

The communication rate in 5G telecommunications has increased significantly compared to previous generations. The rate of telecommunication communication has increased from $2 kb/s$ to $1 Gb/s$, and it is expected to grow a lot in generations beyond the 5G. Communication rate is a very valuable component to evaluate network performance and also has a direct impact on network EE [21], [22], [23]. The use of UAVs in new generation telecommunications has been an important factor in increasing the communication rate. In the proposed scenario of this paper, it is also seen how the rate of telecommunication communication increases with the use of UAVs as BSs [24], [25].

All in all, the application of EH techniques in telecommunication networks has been studied in detail in recent studies. Most of the existing works have investigated EH between uplink users and UAVs. According to the available studies, EH in addition to CR leads to a great improvement in EE and SE. To the best of our knowledge, SWIPT techniques between cellular users and D2D users have not been used in the existing works in the field of using UAVs in wireless networks, and most of the existing works deal with the implementation of energy harvesting links between UAVs and ground users. On the other hand, in this paper, for the first time, SWIPT techniques have been implemented between cellular and D2D users in addition to cognitive radio in wireless networks. The energy efficiency of the network has been investigated and it is shown that an acceptable energy efficiency can be achieved in such a scenario.

\subsection{Motivations and Key Contributions}
The motivation behind this paper is to investigate the potential benefits of utilizing cognitive radio (CR) technology and unmanned aerial vehicle (UAV) in the context of disaster response networks (DRNs). Specifically, we focus on two scenarios: a free disaster response network (free-DRN) and an energy harvesting enabled DRN (EH-enabled DRN). The goal is to assess the impact of these technologies on energy efficiency (EE) and communication rate within the network and compare the performance of the two scenarios.

The Key Contributions can be shown as follows:
\begin{enumerate}
  \item Energy Efficiency Improvement: The paper demonstrates that in the EH-enabled DRN scenario, where device-to-device (D2D) communication links harvest energy from cellular users, there is a significant improvement in both energy efficiency (EE) and communication rate compared to the free-DRN scenario. This finding highlights the potential benefits of energy harvesting techniques in enhancing the performance of DRNs.
  \item Network Performance Analysis: The authors analyze the network performance in both scenarios by considering factors such as user density and UAV specifications. They study how changes in these parameters affect the network's energy efficiency. Additionally, the paper addresses the challenge of managing the time spent on energy harvesting by cellular users and D2D devices in the second scenario.
  \item Optimal UAV Height: The simulation results lead to the determination of an optimal height at which the UAV should fly to ensure a well-functioning network. This finding provides practical guidance for deploying UAVs in DRNs, considering different densities of D2D devices.
  \item Positive Effect of EH-enabled DRN: The numerical and simulation results highlight the advantages of the EH-enabled DRN scenario over the free-DRN scenario. The EH-enabled DRN allows D2D devices to establish efficient communication links, resulting in improved network energy efficiency. Moreover, reducing the duration of energy harvesting and decreasing the density of cellular users can further enhance the network's energy efficiency by up to 3dB.
  \item Coexistence of Uplink Cellular Users and D2D Users: The paper proposes two scenarios that enable communication services in a DRN, where uplink cellular users and D2D users coexist and share frequency bands. The UAV serves as a flying base station (BS) in both scenarios. The second scenario incorporates energy harvesting, allowing D2D users to transfer data by harvesting energy from uplink users, while the first scenario lacks this capability. The paper examines the performance of the DRN by analyzing user rates and overall network energy efficiency.
  \item Impact of User Density: The authors investigate the impact of user density on the communication rate of uplink users and D2D users. They find that increasing D2D user density leads to a decrease in uplink communication rate due to interference from D2D users. However, the behavior of D2D user density differs due to changes in harvested power and interference effects. Managing this trade-off between user density and network performance becomes crucial for future investigations.
\end{enumerate}

Overall, this paper contributes to the understanding of the benefits and challenges associated with integrating CR technology, UAVs, and energy harvesting techniques in disaster response networks. It provides insights into how these technologies can improve energy efficiency, communication rates, and overall network performance, highlighting the potential for more effective and sustainable disaster response strategies.

\subsection{Organization}
The organization of the paper is organized as follows. The system framework is outlined in section II. Then in section III, the formulation of problems is given. Simulation results are illustrated in section IV. Finally, the conclusion of the paper is shown in section V.

\section{Framework Outline}
In both proposed scenarios of this paper, it is assumed that there are $Q$ frequency bands, where $W^q$ indicates the bandwidth of $q^{th}$ frequency band. All users are distributed according to homogeneous Poisson Point Process (PPP) distributions, $\Phi ^c$ with density $\Lambda ^c$ and $\Phi ^{D2D}$ with density $\Lambda ^{D2D}$ respectively for cellular users and D2D users. D2D users harvest energy from cellular users to provide enough energy to establish D2D links. Following that, the UAV communicates with cellular users in an uplink approach which all uplink communications are orthogonal, so cellular users do no interference for each others.

\subsection{Communication Model}
The main difference between scenarios is that there is no C2D EH link in the first scenario. Following that, the transfer power of D2D transmitters is the main challenge of the DRN in the first scenario. But, C2D EH links are established between cellular users and transmitters of D2D pairs. The formulations presented in the rest of the paper all apply to both scenarios. Only in the first scenario, D2D users have limited sending power, but in the second scenario, their sending power is enhanced by cellular users.

There are two different channels in two given scenarios:
\begin{itemize}
  \item Ground to Ground (G2G) channel: The Rayleigh fading (large-scale path-loss and the small-scale fading) channel model that is established between the transmitter and receiver of D2D pairs, as well as it is established between cellular users and transmitters of D2D pairs in the second scenario. There is a an C2D EH link between cellular users and transmitters of D2D pairs. The power of the received energy signal in the transmitter of the $b^{th}$ D2D user from the $a^{th}$ cellular user on the $q^{th}$ frequency band can be defined as: $P^{E,b,q}=P^{T,c,q}f^{a,b}l_{min}^{-\alpha _g}$, where $P^{T,c,q}$ denotes the transfer power of each cellular user on the $q^{th}$ frequency band, $f^{a,b}$ indicates the norm of the gain of the G2G channel which follows an independent exponential distribution with unit mean, $l_{min}$ shows the distance between the transmitter of the $b^{th}$ D2D pair and the closest cellular user that is $a^{th}$ cellular user, and $\alpha _g$ is the path-loss exponent of G2G channel .Furthermore, the power of the received signal in the receiver of the $b^{th}$ D2D pair from its transmitter on the $q^{th}$ frequency band is given as: $P^{b,q}=P^{T,d,q}f^{b}l_{D2D}^{-\alpha _g}$, where $P^{T,d,q}$ shows the transfer power of each D2D transmitter on the $q^{th}$ frequency band that will be provided by cellular users, $f^{b}$ denotes the norm of the gain of the G2G channel and $l_{D2D}$ is the distance between the transmitter and the receiver of D2D pairs.
  \item Ground to Air (G2A) Channel: This uplink channel is established between terrestrial cellular users and the UAV which all uplink signals are composed of the combination of line-of-sight (LoS), non-LOS (NLoS), and other reflected components. The probability of each component which is a function of environmental parameters; $P^{los} + P^{nlos} = 1$, and then according to Fig. 2 it is calculated for each cellular user as: $P^{los}=\frac{1}{1+C exp(-B [\Theta -C])}$ ,where $\Theta = \frac{180}{\pi} sin^{-1}(\frac{H}{D_{a}})$. B and C denote constant values of environment. Following that the power of the received signal at the UAV from $a^{th}$ cellular user on the $q^{th}$ frequency band can be defined as: $P^{a,q}=(P^{los} +\eta P^{nlos})P^{T,c}D_{a}^{-\alpha _a}$, where $\eta$ is the extra attenuation factor of nlos links and $\alpha _a$ indicates the path-loss exponent of G2A channel.
\end{itemize}
\begin{figure}[!t]
\centerline{\includegraphics[width=\columnwidth]{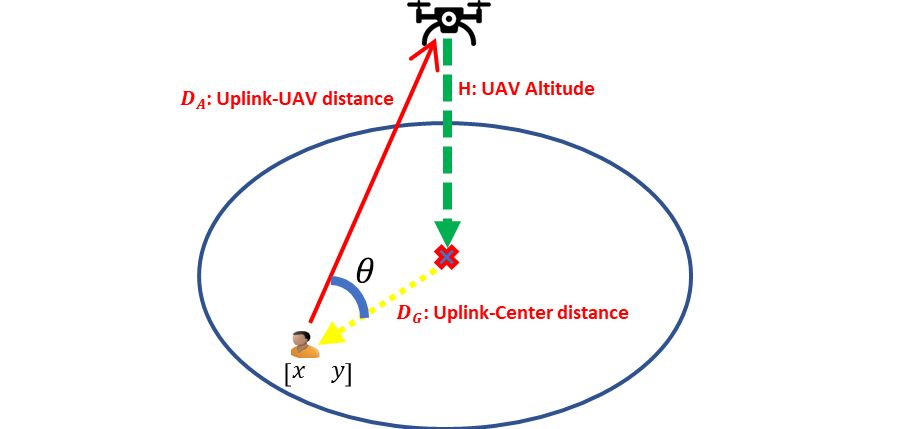}}
\caption{plos and pnlos calculation according to the elevation angle($\Theta$)}
\label{fig2}
\end{figure}

In both scenarios, D2D users transmissions interference for each other and also for the transmission of uplink users in the UAV. Moreover, uplink users interference for D2D users. The power of the interference signal of other D2D users for the receiver of $b^{th}$ D2D pair on the $q^{th}$ frequency band can be defined as follows:
\begin{equation}
    I_{d}^{b,q}=\sum_{i\epsilon \Lambda ^{D2D}, i\ne b}P^{T,d}f^{i,b}l_{i,b}^{-\alpha _g},
\end{equation}
where $f^{i,b}$ indicates the norm of the gain of the G2G channel between the transmitter of $i^{th}$ D2D pair and the receiver of $b^{th}$ D2D pair and $l_{i,b}$ denotes the distance between mentioned users.

Furthermore, cellular users interference for the receiver of the $b^{th}$ D2D pair on the $q^{th}$ frequency band that can be calculated as:
\begin{equation}
    I_{c}^{b,q}=\sum_{j\epsilon \Lambda ^{c}}P^{T,c}f^{j,b}l_{j,b}^{-\alpha _g},
\end{equation}
where $f^{j,b}$ indicates the norm of the gain of the G2G channel between $j_{th}$ cellular user and the receiver of $b_{th}$ D2D pair and $l_{j,b}$ is the distance between these users.

On the other hand, D2D communications play the role of the interference signal for received uplink signals in the UAV. The interference of D2D transmitters for uplink signals on the $q^{th}$ frequency band can be defined as:
\begin{equation}
    I^{c,q}=\sum_{i\epsilon \Lambda ^{c}}P^{T,d}f^{i,u}l_{i,u}^{-\alpha _a},
\end{equation}
where $f^{i,u}$ denotes the norm of the gain of the G2A channel between the transmitter of $i_{th}$ D2D pair and the UAV and $l_{i,u}$ indicates the distance between this user and the UAV.

\subsection{Energy Harvesting Model}
power splitting energy harvesting architecture has been used to harvest energy in the proposed scenario of this paper.Power splitting energy harvesting mode, also known as power splitting energy harvesting (PSEH), is a technique used in energy harvesting systems to maximize the utilization of harvested energy. In PSEH, the harvested energy is split into two or more branches, each serving a different purpose or load. The basic concept of power splitting energy harvesting mode involves dividing the harvested energy into two components: one for immediate power usage and the other for energy storage. This division allows for simultaneous power delivery to a load and energy accumulation for future use. By employing power splitting energy harvesting mode, the system achieves a balance between powering the load in real-time and accumulating surplus energy for future use. This mode increases the overall system efficiency and improves the reliability of energy harvesting systems, especially in scenarios where the energy source is intermittent or unpredictable.

According to PSEH architecture, $\delta^{PSEH} P^{E,b,q}$ is directed to the data decoding unit and $(1-\delta^{PSEH}) P^{E,b,q}$ to the energy harvesting unit. A nonlinear energy harvesting model is considered in this paper in which the harvested power of the transmitter of $b^{th}$ D2D pair from $a^{th}$ cellular user on the $q^{th}$ frequency band in the second scenario can be given as follows [26], [27]:
\begin{equation}
    p^{b,q} = alpha^1 ((1-\delta^{PSEH}) P^{E,b,q})^2 + alpha^2 ((1-\delta^{PSEH}) P^{E,b,q}) + alpha^3,
\end{equation}
where $alpha^1, alpha^2, alpha^3 \epsilon R $ are parameters of the PSEH model and $0 /ge \delta^{PSEH} \le 1$.

Harvested energy from cellular users will be spent on D2D communications in the second scenario, so the transfer power of D2D users is enhanced. It is assumed that cellular users spend a certain amount of time for energy links that is indicated by $T^{th,c}$. Hence, the energy that the transmitter of $b^{th}$ D2D pair harvests from $a^{th}$ cellular user on the $q^{th}$ frequency band in the second scenario is calculated as follows:
\begin{equation}
    E^{b,q}=T^{th,c}p^{b,q}.
\end{equation}

Then, it is assumed that all D2D transmitters spend a certain amount of time to establish D2D link to their receivers that is defined as $T^{th,d}$. So, the transfer power of D2D transmitters in the $q^{th}$ frequency band can be calculated as:
\begin{equation}
    P^{T,d,q}=\frac{E^{b,q}}{T^{th,d}}.
\end{equation}

\section{Rate and EE Evaluation}
In this section, the SINR of uplink and D2D communications will be defined and then the rate of communication of each user will be calculated and finally the EE of the DRN will be given. As uplink communications are orthogonal, so the SINR of $a_{th}$ uplink user on the $q^{th}$ frequency band can be given as:
\begin{equation}
    \gamma^{a,q} =\frac{P^{a}}{I^{c}+n^{a}},
\end{equation}
where $n^{c}$ denotes the power of Additive White Gaussian Noise (AWGN) received at the UAV through G2A channel.

The interference term for D2D received signals is more powerful than its term for uplink communications, it is composed of D2D signals and uplink signals. The SINR for the receiver of $b^{th}$ D2D pair on the $q^{th}$ frequency band is given as:
\begin{equation}
    \gamma _{d}^{b,q}=\frac{P^{b}}{I_{d}^{b}+I_{c}^{b}+n^{g}},
\end{equation}
where $n^{g}$ denotes the power of AWGN received at the receiver of D2D pairs through G2G channel.

The rate of the $a_{th}$ uplink user served by the UAV on the $q^{th}$ frequency band is given as follows:
\begin{equation}
    \zeta^{a,q}=W_q log_2(1+\gamma _{d}^{b,q}).
\end{equation}

Following that, the rate of the receiver of the $b_{th}$ D2D pair on the $q^{th}$ frequency band is defined as:
\begin{equation}
    \zeta_{d}^{b,q}=W_q log_2(1+\gamma^{a,q}).
\end{equation}

Finally, the EE of whole uplink communications and D2D communications in the $q^{th}$ frequency band are calculated as follows:
\begin{equation}
    EE^{q}=\frac{(\sum_{i\epsilon \Lambda^{D2D}}\zeta_{d}^{b,q})+(\sum_{j\epsilon \Lambda ^{c}}\zeta^{a,q})}{(T^{th,a}+T^{th,c})\sum_{j\epsilon \Lambda ^{c}}P^{T,c,q}}.
\end{equation}

It is assumed that each uplink user spends $T^{th,a}$ communicating to the UAV.
Based on (10), the EE of on all frequency bands which can be called as total EE of the whole DRN can be calculated as given:
\begin{equation}
    EE^{total}=\frac{\sum_{q=1}^{Q}({(\sum_{i\epsilon \Lambda ^{D2D}}\zeta_{d}^{b,q})+(\sum_{j\epsilon \Lambda ^{c}}\zeta^{a,q})})}{\sum_{q=1}^{Q}((T^{th,a}+T^{th,c})\sum_{j\epsilon \Lambda ^{c}}P^{T,c,q})}.
\end{equation}

The interaction between the EE and the parameters of the DRN and some features of the UAV are shown in (10) and (11), and some valuable notes can be illustrated based on these equations. First of all, as the transfer power of cellular users increases, harvested energy by D2D users grows, and accordingly the transfer power of D2D users and then the SINR of D2D users will go up. On the other hand, there is a trade-off between $T^{th,d}$, $T^{th,c}$ and the density of users which have serious effect on the EE of whole DRN, and it will be examined completely in next section. At the end, the SINR and the rate of D2D users do not depend on the parameters of the UAV.

Note that, the transfer power of D2D users are fixed in the first scenario. But, the transfer power of the transmitter of D2D users is enhanced in the second scenario and (4) shows the extra energy that D2D transmitter access in the second scenario. In the second scenario, compared to the first scenario, the energy of cellular users is optimally consumed, and the transmission power of D2D users is also strengthened. As a result, according to (9) and (11), network EE and D2D communication rate in the second scenario have a much better value than the first scenario, and these results are also shown in the simulation section.

\section{Simulation Results}
In this section, the advantage of the second scenario compared to the first scenario is proved according to the simulation results. The performance of EE of both the uplink communications and D2D communications are studied, using both mathematical results derived in sections II and III, as well as simulation results. All essential simulation parameters are listed in Table 1. The effect of various parameters of the DRN and the UAV such as users' density, transfer power of uplink users, and the altitude of the UAV on the EE of whole DRN will be investigated.
\begin{table}
\centering
\caption{Simulation Parameters}
\label{table}
\setlength{\tabcolsep}{25pt}
    \begin{tabular}{|p{25pt}|p{75pt}|}
        \hline
        Parameter&
        Value \\
        \hline
        $\alpha _a$&
        2 \\
        \hline
        $\alpha _g$&
        4 \\
        \hline
        $\eta$&
        20 dB \\
        \hline
        B&
        0.136 \\
        \hline
        C&
        11.95 \\
        \hline
        Q&
        5 \\
        \hline
        $n^a$&
        -120 dBm \\
        \hline
        $n^g$&
        -120 dBm \\
        \hline
        $W^q$&
        $10^8 MHz$ \\
        \hline
        $alpha^1$&
        -0.116 \\
        \hline
        $aalpha^2$&
        $0.6574$ \\
        \hline
        $alpha^3$&
        $-6.549 \times 10^{-7}$ \\
        \hline
        $\delta^{PSEH}$&
        0.1 \\
        \hline
    \end{tabular}
\end{table}

Fig. 3 illustrates the effect of the density of D2D users on the EE of whole DRN in $q^{th}$ frequency band for both scenarios. It is assumed that $\Lambda ^c=3*10^{-4}\ users/m^2$, $l_{min}=10\ meter (m)$, $l_{D2D}=10\ m$, $T^{th,c}=0.05\ second (s)$, $T^{th,d}=0.05\ s$, $T^{th,a}=0.05\ s$ and $H=700\ m$. Increase in Density of D2D users leads to an increase in interferences that D2D users create for uplink and D2D communications, but at first when the density of D2D users is low, the effect of the desired signal that D2D users send to their receivers is lower than the effect of the interference. Then, with more increase in the density, the effect of D2D users' interferences becomes more powerful than desired D2D signals. As expected, the network performance in the second scenario is much better than the first scenario. So that even in some values of D2D user densities, $3 \times 10^{-4} pairs/{m^2}$, the EE in the second scenario is $0.5 GBps/J$ higher than the first scenario. Moreover, there are $\% 2$ of improvement in terms of EE of whole DRN when the density of D2D users increased from $\Lambda ^{D2D}= 6\times 10^-4 pairs/{m^2}$ to $\Lambda ^{D2D}= 10^-3 pairs/{m^2}$ in both scenarios.
\begin{figure}[h]
    \label{Fig5}
    \centering
    \includegraphics[width=3.5in,trim={0 6cm 0 6cm},clip]{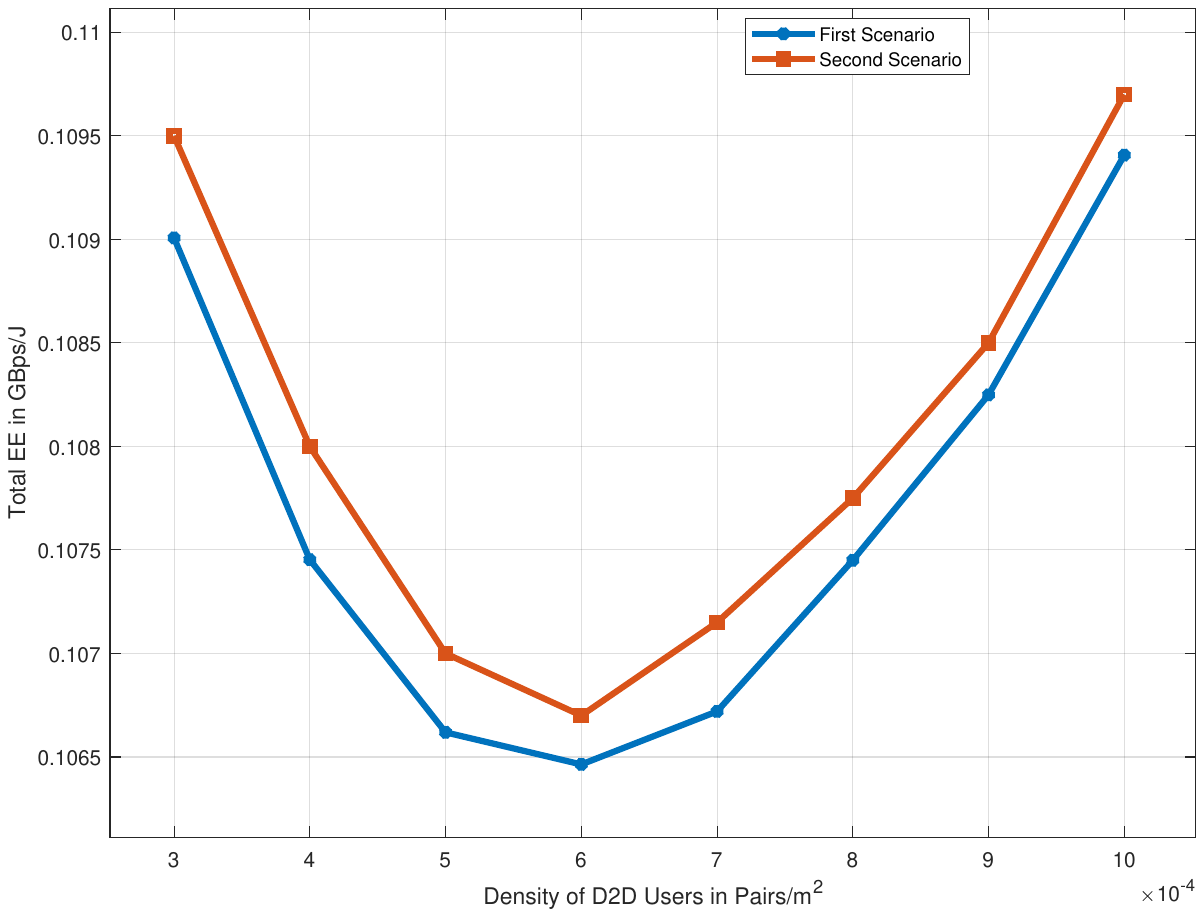}
\caption{Total EE of DRN vs. Density of D2D users in both Scenarios}
\end{figure}

The impact of changes in the density of cellular users is shown in Fig. 4. Simulation parameters are like the previous one. The interference that uplink signals have on D2D signals and also with the increase in the density of uplink users, more energy is consumed. Hence, increase in the density of uplink users has a negative effect on EE of whole DRN. Furthermore, when the altitude of the UAV decreases, the power of useful signals will increase, so the EE will grow. As expected from theoretical results and shown in this figure, the applying the EH technique in the second scenario is resulted in notable improvement in the EE of the DRN. For instance, in density of $8 \times 10^{-3} users/{m^2}$ and $H=500 m$, the EE in the second scenario is $100 GBps/J$ higher than the first scenario. It is notable that if the density of uplink users decreases from $\Lambda ^{c}= 10^-3 users/{m^2}$ to $\Lambda ^{c}= 3\times10^-4 users/{m^2}$, there will be $\% 18.5$ of improvement in terms of EE of whole DRN in both scenarios.
\begin{figure}[h]
    \label{Fig6}
    \centering
    \includegraphics[width=3.5in,trim={0 6cm 0 6cm},clip]{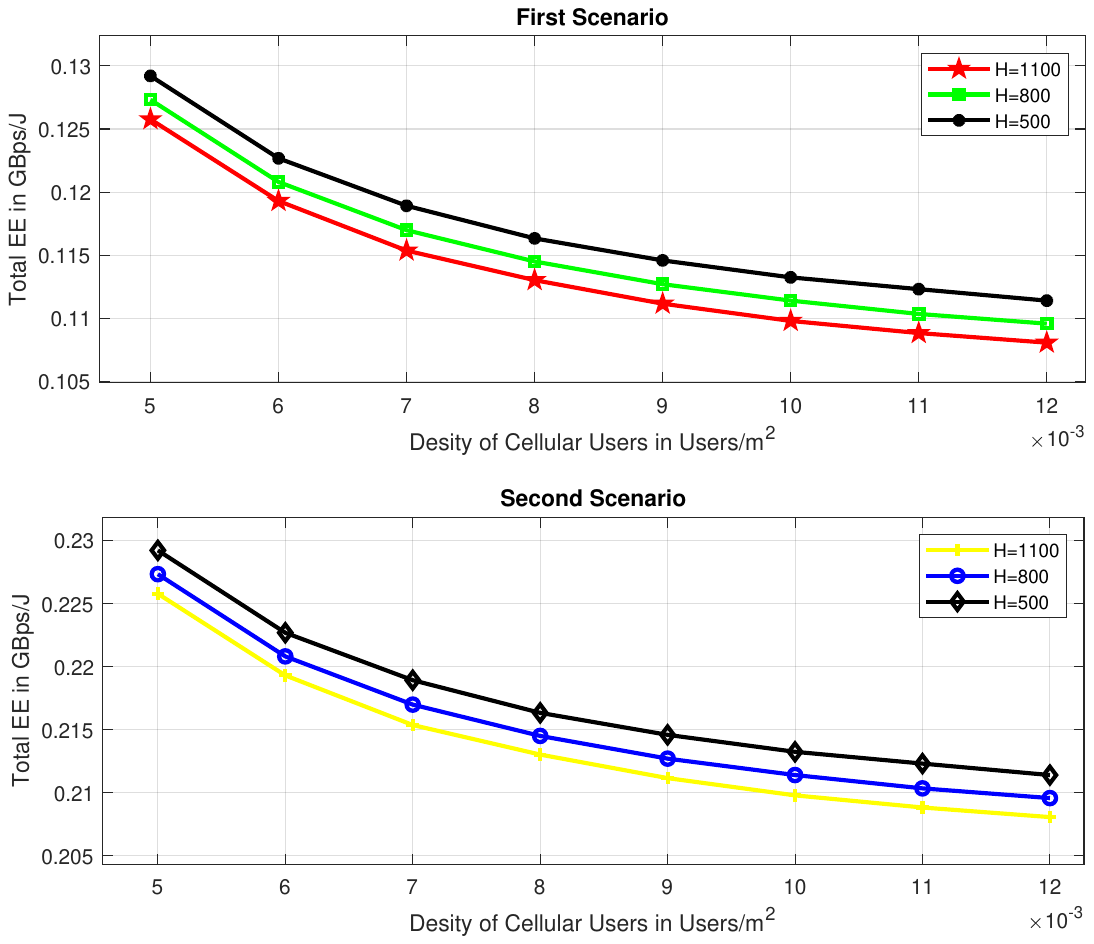}
\caption{Total EE of DRN vs. Density of Uplink users and UAV altitude}
\end{figure}

Fig. 5 indicates the relationship between the altitude of the UAV and density of D2D users to achieve a certain amount of EE. Simulation parameters are like previous ones. As expected, when the density of D2D users increases, the UAV forces to hover in lower altitude to reach to a special amount of EE. As a result, an optimal altitude for the UAV is calculated. As other figures, this figures proves the effectiveness of using the EH technique in the second scenario in compared to the first scenario. When the density of D2D users is $\Lambda ^{D2D}= 8\times 10^-4 pairs/{m^2}$, the UAV has to fly at an altitude 100 meters lower than in the second scenario in order to meet the network requirement. According to this figure, by increasing the density of D2D users from $\Lambda ^{D2D}= 2\times 10^-4 pairs/{m^2}$ to $\Lambda ^{D2D}= 5\times 10^-4 pairs/{m^2}$, the UAV will be forced to halve its height to meet the EE threshold of DRN in both scenarios.
\begin{figure}[h]
    \label{Fig5}
    \centering
    \includegraphics[width=3.5in,trim={0 6cm 0 6cm},clip]{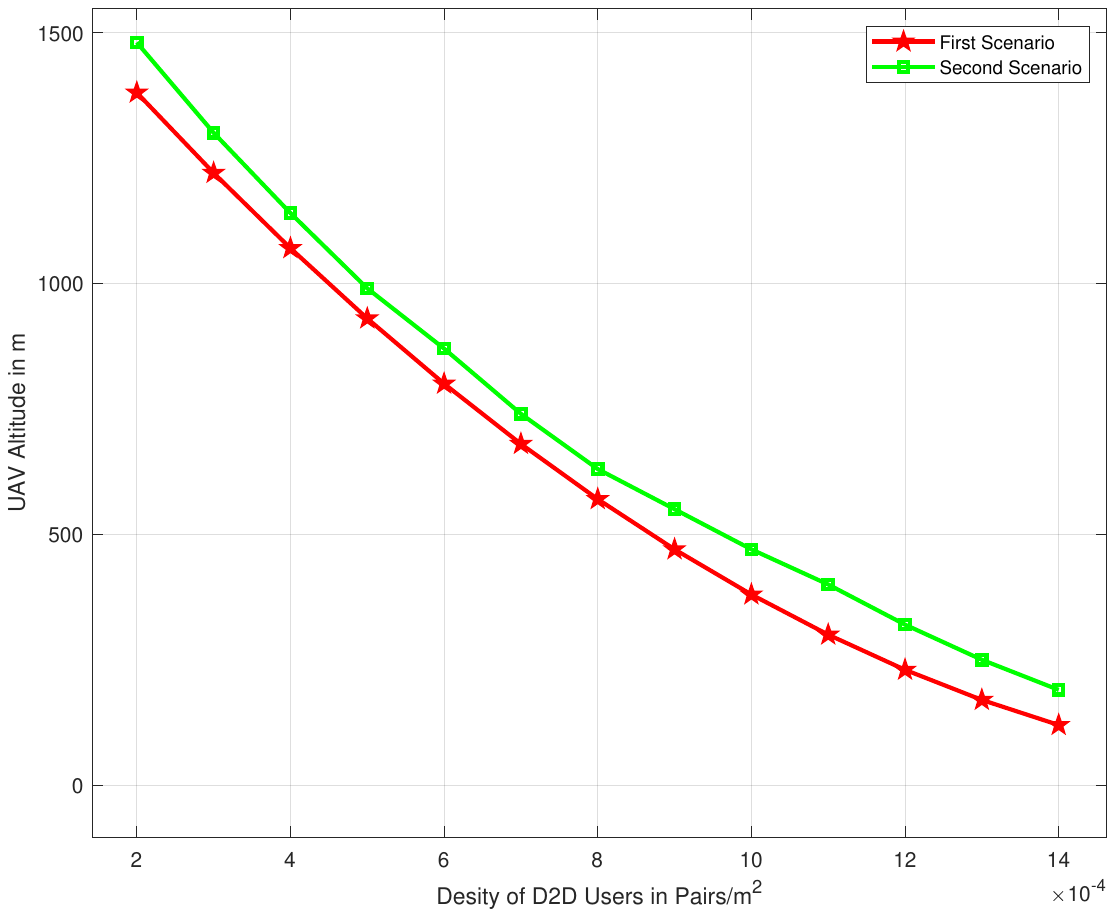}
\caption{UAV Altitude vs. Density of D2D users}
\end{figure}

According to Fig. 6, as the time spent by cellular users to transfer energy to D2D users increases in the second scenario, the EE of the network decreases. Simulation parameters are like previous ones. If cellular users spend a lot of time transferring energy, the amount of energy increases, which is inversely related to EE. In addition, when the distance between the closest cellular user and the transmitter of a D2D pair, $l_{min}$, declines, D2D users harvest much more energy, so the EE declines. When the distance between the closest cellular user and the transmitter of a D2D pair gets a quarter, there will be $\% 50$ of improvement in terms total EE of DRN.
\begin{figure}[h]
    \label{Fig6}
    \centering
    \includegraphics[width=3.5in,trim={0 6cm 0 6cm},clip]{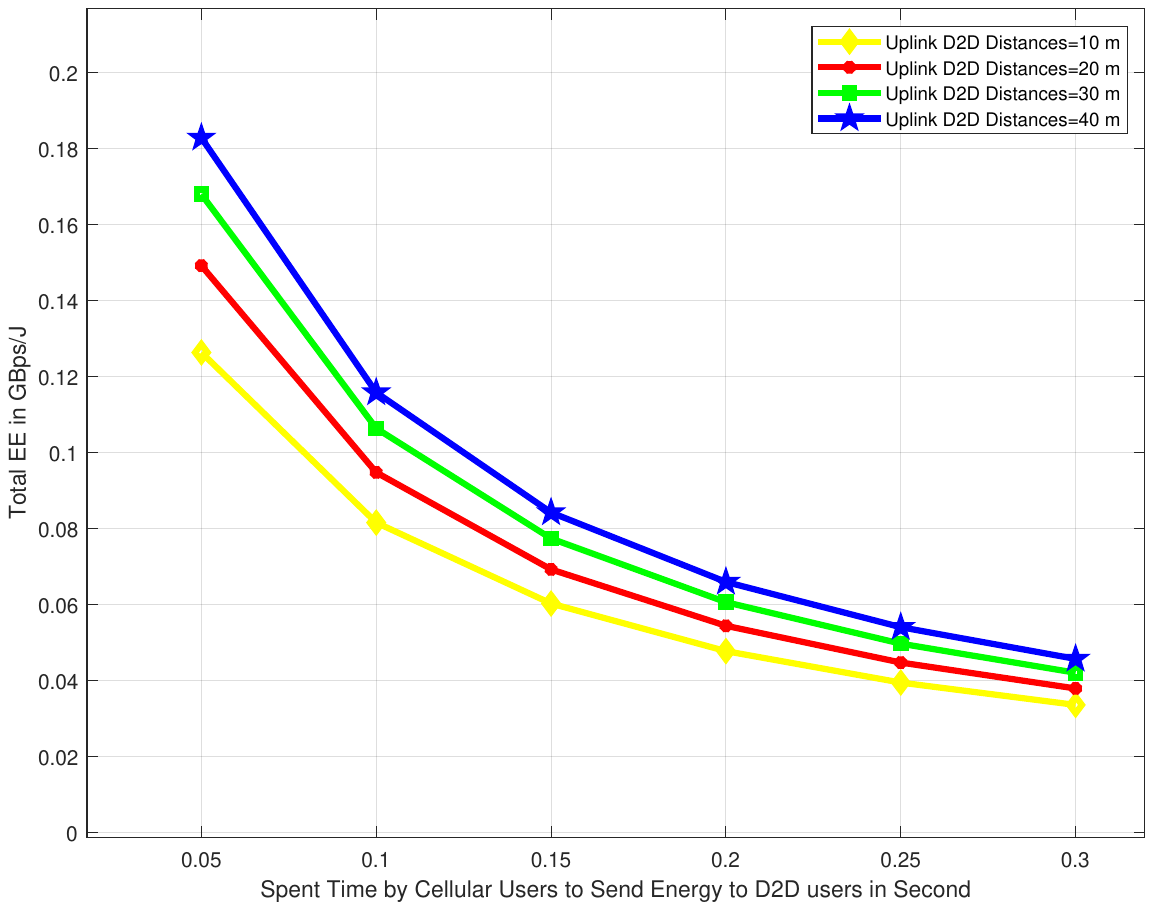}
\caption{Total EE of DRN vs. Spent Time by Cellular Users to Send Energy to D2D users}
\end{figure}

As shown in Fig. 7, the behavior of the data transmission rate in uplink communication and D2D communication are completely different with the change of user density in both scenarios. As the density of uplink users increases, the rate of D2D communication increases, while as the density of D2D users increases, the rate of uplink communication decreases. As uplink users increase, D2D users have more energy resources, so their transmission power is improved and they achieve better rates. On the other hand, when the density of D2D users increases, the interference of these users to the uplink users increases, resulting in a decrease in the uplink transmission rate. In addition, the rate of uplink communication does not depend on the EH technique, so it is the same in both scenarios. In contrary, due the applying the EH links between cellular users and D2D users, the rate of D2D users is much better in the second scenario rather than the first scenario. According to Fig. 7, when the density of D2D users becomes one third in both scenarios, the uplink communication rate increases by 1 dB, which is a good achievement.
\begin{figure}[h]
    \label{Fig7}
    \centering
    \includegraphics[width=3.5in,trim={0 6cm 0 6cm},clip]{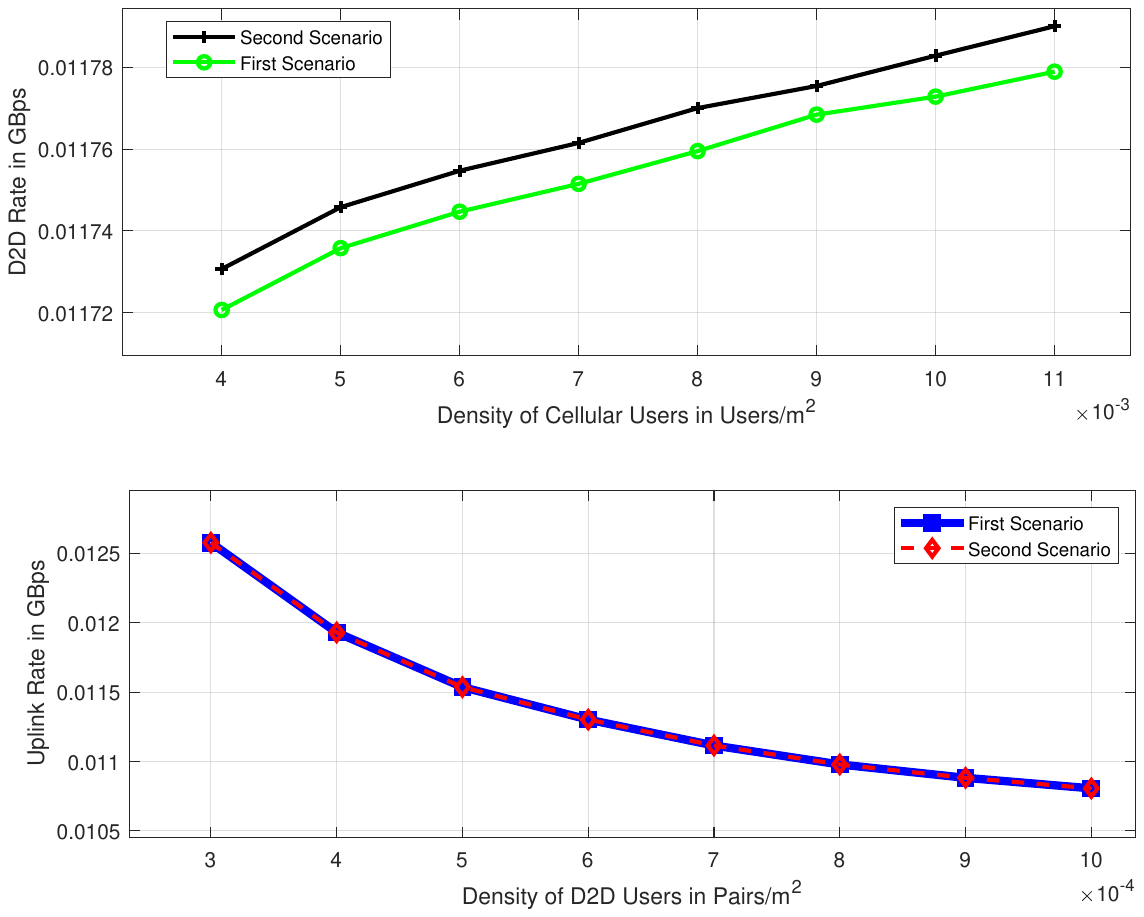}
\caption{Rate of D2D and Uplink Communications vs. Users' Density}
\end{figure}

In conclusion, the simulation results demonstrate the impact of density variations of D2D users and cellular users on the energy efficiency (EE) and data transmission rates in the considered scenarios. It is observed that increasing the density of D2D users leads to higher interferences and a decrease in the desired signal power, resulting in a decline in EE. However, in the second scenario, where energy harvesting (EH) technique is employed, notable improvements in EE are achieved compared to the first scenario. The EH technique helps in mitigating interferences and increasing the transmission power of D2D users, leading to enhanced EE and data rates. Furthermore, the optimal altitude for the UAV is calculated based on the density of D2D users, and it is found that the UAV needs to hover at a lower altitude in the second scenario to meet the network requirements. Additionally, the time spent by cellular users in energy transfer and the distance between cellular users and D2D transmitters impact the overall EE, with longer energy transfer times and shorter distances resulting in lower EE. Overall, the simulation results highlight the effectiveness of the EH technique and the influence of user densities on the performance of the network.

\section{Conclusion}
In conclusion, this paper proposed two scenarios for providing communication services to a Device-to-Device Relay Network (DRN), where uplink cellular users and D2D users coexist and share frequency bands. In both scenarios, an Unmanned Aerial Vehicle (UAV) acts as a flying base station. The first scenario focused on direct D2D communication without energy harvesting (EH) capabilities. In the second scenario, D2D users harvested energy from uplink users to facilitate data transfer to their receivers. By analyzing the performance of the DRN in terms of user rate and energy efficiency (EE), several key findings were observed. Firstly, an inverse relationship between the density of uplink users and the EE of the DRN was identified. Higher uplink user density resulted in reduced EE due to increased energy consumption. Conversely, the behavior of D2D user density was different due to changes in harvested power and interferences caused by D2D users and uplink users. Although increasing harvested energy improved D2D communication, it also led to increased energy consumption and, consequently, decreased EE of the network. The simulation results revealed that halving the EH time for uplink users in the second scenario resulted in a doubling of the EE of the entire DRN. Furthermore, the density of users had varying effects on the uplink communication rate and D2D communication rate. Increasing D2D user density led to a downward trend in the uplink communication rate, while the D2D communication rate also showed a similar downward trend. This disparity was primarily due to the interference caused by D2D users on cellular users in both scenarios and the utilization of EH between cellular users and D2D users only in the second scenario. Overall, the second scenario, incorporating the EH technique, provided a more favorable situation for users. However, finding an optimal trade-off between the amount of harvested energy from cellular users and the EE of the network remains a significant challenge for scenarios like the second scenario. This challenge presents an avenue for future investigations and further study.

\end{document}